# Why a Global Time is Needed in a Dependable SoS

Hermann Kopetz
Institute für Technische Informatik
Vienna University of Technology
1040 Vienna, Austria
e-mail: H.Kopetz@gmail.com

*Abstract*—A *system-of-systems* (SoS) is a large information processing system formed by the integration of autonomous computer systems (called constituent systems, CS), physical machines and humans for the purpose of providing new synergistic services and/or more efficient economic processes. In a number of applications, e.g robotics, the autonomous CSs must coordinate their actions in the temporal domain to realize the desired objectives. In this paper we argue that the introduction of a proper global physical time establishes a shared view about the progress of physical time and helps to realize the temporal coordination of the autonomous CSs. The available global time can also be used to simplify the solution of many challenging problems within the SoS, such as distributed resource allocation, and helps to improve the dependability and fault-tolerance of the SoS.

## I. INTRODUCTION

A *System-of-System (SoS)* is characterized by the cooperation of autonomous computer systems, called *constituent systems (CS),* in order to realize composite functions that cannot be provided by any of the CSs in isolation. Many of the CSs that form an SoS are cyber-physical systems (CPS), where the autonomous computer system manipulates a process in the physical world.

Whenever a physical process is controlled by a computer systems two different notions of time must be carefully distinguished: (i) the *physical time* that is an independent variable in the laws of physics describing the behavior of the physical process as a function of time and (ii) the *logical time* that characterizes some aspects of the behavior of the computer system that unfolds as a consequence of the execution of the stored programs.

In Newtonian physics, the *physical time* is considered an independent variable with a dense domain that progresses from the past to the future—the *timeline* that is often called *the arrow of time*. A *cut* of the timeline is called an *instant* and a *section* of the timeline is called a *duration*. A happening at an instant is called an *event*. The position of an event on the timeline can be recorded by *time-stamping the event* relative to the state of a clock, i.e., by assigning the state of a physical clock at the instant of event occurrence to the event.

The need for a worldwide physical time standard that can be generated in a laboratory gave birth to the International Atomic Time (TAI). TAI defines the second as the duration of 9,192,631,770 periods of the radiation of a specified transition of the cesium atom 133. The epoch of TAI starts on January 1, 1958 at 00:00 h Greenwich Mean Time (GMT). The time base of the global positioning system GPS is based on TAI with the epoch starting on January 6, 1980 at 00:00 h [TAI13].

A *logical time* is introduced inside a computer system to establish the causal order among computational events and to support the reasoning about properties of sequential and concurrent computational processes [Ray00]. There is no metric associated with logical time. Due to the complex architecture of todays CPUs and the unpredictable delay of many communication systems, it is difficult to establish the exact physical duration between significant computational events, even if the frequency of the physical oscillator that drives the CPU and the exact program that produces the computational events are known.

Whereas in a monolithic computer a *common clock* for *time-stamping events* can be derived directly from the signals of the central physical oscillator, no such *common clock* exists in an SoS. Each autonomous CS has its own oscillator that swings freely and is uncoordinated with respect to the oscillation of the oscillator in any other autonomous CS. It is thus not possible to measure the duration between events that occur in the physical environment of different CSs if no global notion of time of adequate precision is shared by all CSs of the SoS. Given that such a global SoS time is available, this global time can be used to radically simplify the solution of many other temporal coordination problems in an SoS.

It is the objective of this paper to elaborate on the manifold uses of a global time in the solution of temporal coordination problems in an SoS. The paper starts with a short characterization of an SoS in Section two. Section three discusses the required properties of a suitable global time base and shows how such a time-base can be established in an SoS. Section four, the main Section of the paper, introduces a number of temporal coordination problems in an SoS and shows how the availability of a global time can support the solution process. The paper terminates with a conclusion in Section five.

## II. SYSTEM-OF-SYSTEMS

The domain of *Systems-of-Systems (SoS)* is a relatively new field of computer science that is concerned with the design and operation of large information processing systems that are composed of existing or new *autonomous constituent computer systems (CS), physical machines and humans* [Jam09]. It is assumed that the integration of the CSs will improve current economic processes and provide new synergistic services. The integration of CSs into SoSs is already happening on a wide scale. It is achieved by the message-based exchange of information processed and stored in the diverse CSs.

Table 1 characterizes a SoS by listing some of the distinguishing properties of a SoS compared to those of a classic monolithic system [Kop13, adapted from Mai98]. If we look at this table we see that the listed characteristics of an SoS violate many of the fundamental assumptions that are taken for granted in the established monolithic system design process. For example, there is no fixed specification, coordinated evolution, or final acceptance test of an SoS.

| Characteristic | Old-Monolithic | New-SoS |
| --- | --- | --- |
| Scope of System | Fixed (known) | Not known |
| Requirements and Spec. | Fixed | Changing |
| Control | Central | Autonomous |
| Evolution | Version control | Uncoordinated |
| Testing | Test phases | Continuous |
| Implementation Technology | Given and fixed | Unknown |
| Faults (Physical, Design) | Exceptional | Normal |
| Emergence | Insignificant | Important |
| System development | Waterfall model | ??? |

**Tab. 1**: Monolithic System versus a System of Systems [adapted from Mai98].

The most differentiating characteristics of a SoS relative to a monolithic system are the *autonomy* of the CSs, the *uncoordinated evolution* of the CSs, the facts that *testing is a continuous activity* that must also be performed during the operational phase of a SoS [Dah10] and that the occurrence of faults must be considered *normal* in an SoS An SoS design must thus provide proper mechanisms for fault-containment, error detection and failure mitigation.

Maximum autonomy of the CSs of an SoS can be achieved if the services provided at the interfaces between the CSs are precisely specified in the domains of *functionality*, *value and time* and no further assumptions need to be made about the internal operations of a CS. The temporal specification of services requires a concept of time that is shared among the interfacing CSs. Since in the Newtonian model of the world the progression of the physical time is *independent* and *omnipresent*, a global *physical time base* establishes a universally available temporal framework that can be used at every interface of a CS to specify temporal conditions without any reference to CS internal implementation choices.

## III. SoS GLOBAL TIME

### A. Properties of the SoS Global Time Base

Before establishing the properties of any SoS global time base we must take account an impossibility results that cannot be overcome by *any* implementation: *It is impossible to precisely synchronize the clocks in a distributed computer system*. We call the maximum temporal difference in the states of any two clocks of an ensemble, measured by an omniscient reference clock the achieved *precision* of the ensemble during the *interval of discourse*. The precision determines the granularity of a reasonable discrete time base. Due to the *precision error* and the *discreteness of the digital time base*, a measurement error in the timestamps of events is unavoidable. This measurement error can lead to inconsistencies about the perceived and recorded temporal order of events.

In order to avoid these inconsistencies, the notions of a *sparse global time base* and of *sparse events* has been introduced [Kop11, p. 62]. In the sparse time model, the timeline is partitioned into an infinite sequence of *permitted* and *forbidden* intervals. The duration of the permitted intervals and of the forbidden intervals is determined by the achieved *precision* of the ensemble. An event occurring in a permitted interval is called a *sparse event* and the corresponding instant of event occurrence a *sparse instant*. Two sparse events are considered to happen simultaneously if they occur during the same permitted interval, otherwise they are considered temporally ordered. Events that are in the sphere of control of a CS (e.g., the instant of starting to send a message) should be *sparse events* that occur during a permitted interval. They can then be consistently ordered in the whole SoS.

Events that are outside the sphere of control of a CS, e.g, events that happen in the physical environment during a forbidden interval, must be assigned to a permitted interval by an agreement protocol that is executed by all involved CSs. The transformation of an event occurring in the forbidden interval to a permitted interval brings about *temporal consistency* at the *expense of temporal fidelity*. We feel that in most application temporal consistency is more important than utmost temporal fidelity (which can be improved by a better precision, if needed).

### B. Establishment of the SoS Global Time Base

An SoS global time base can be established by *internal clock synchronization* among the CSs of the SoS, or by *external clock synchronization* that brings all clocks into agreement with a trusted external time server that distributes the world standard TAI. The *achievable precision* depends on the quality of the local oscillators, the frequency of resynchronization and the jitter of the communication system.

Internal clock synchronization can be accomplished by the deployment of an appropriate communication protocol. For example, the time-triggered protocol TTP provides an integrated fault-tolerant internal clock synchronization service to all nodes using the protocol [Kop93]. The TTEthernet Standard [Ste08] includes a fault-tolerant internal clock synchronization protocol that is realized on top of standard Ethernet.

In an SoS context it is proposed to synchronize all clocks also by external synchronization to a trusted external time service, such as the time distributed globally by navigation satellite systems (e.g., GPS or Galileo). These timing signals support the establishment of a worldwide SoS global time base with an accuracy of better than 100 nsec [Hof07] The granularity of the respective *global sparse time* will then be better than 1 μsec.

In an SoS where a high synchronization accuracy and a high dependability of the global time-base are required a combination of external and internal clock synchronization will bring the best results. The external synchronization provides for *the long-term accuracy*, while the internal synchronization helps to achieve *a high availability* of the global time in case the external synchronization fails intermittently.

IV. USING THE SOS GLOBAL TIME

In this Section we elaborate on different use cases that can take advantage of an available SoS global time.

*A. Global Timestamps*

In many applications the duration between events that occur in the environment of the different CSs of an SoS must be determined. If a global time-stamp is assigned to every significant event, then the duration between any two significant events occurring at any place within the whole SoS can be calculated easily.

Take the example of measuring the duration it takes for a skier to complete a downhill competition. If a global time of adequate precision is available at every node, the global timestamp of the start event and the global timestamp of the terminating event can be recorded instantaneously. The calculation and display of the time difference can be performed without any pressing real-time requirement—the availability of a global time base takes the *real-time pressure* out of most of the system.

Take the temporal validity of real-time data as another example. An observation of a dynamic entity, e.g., the state of traffic light, e.g., *green*, can only be used for control purposes within a validity interval that depends on the dynamics of the entity (the traffic light). If the observation of the environment is performed by a CS that is different from the CS that uses the observation, then, based on the timestamp of the observation, the user can determine if it safe to use a given observation at a particular moment.

Another most important use of global time relates to the specification of the temporal properties of the interfaces between the CSs. A CS interface specification should be self-contained and inform about the *functionality* and the *state* of the respective interface model. Since the interface state depends on the progression of time, a time value must be part of the interface specification. If a *global time value* is contained in an interface specification, it can be correctly interpreted by all CSs that access the interface, since at any *sparse instant* the (synchronized) global time is the same in all CSs of the SoS.

*B. Synchronization of Input Actions*

An SoS global time can be used to synchronize input actions that occur in different CSs of the SoS to arrive at a common view of the state of the environment at a given instant.

In a smart grid application that extends over a wide geographic region, the state of the grid at a chosen instant can only be established in a central control room if all sensors in the different CSs observe the state of the grid at about the same instant. This is achieved by synchronizing the PMU (Phase measurement units) with GPS time [Bak11].

On August 14, 2003 a major power blackout occurred in parts of the US and Canada. In the final report [USC04] about this blackout it is stated on p. 162: *A valuable lesson from the August 14 blackout is the importance of having time-synchronized system data recorders. The Task Force's investigators labored over thousand of data items to determine the sequence of events, much like putting together small pieces of a very large puzzle. That process would have been significantly faster and easier if there had been wider use of synchronized data recording devices.*

In a financial transaction system, different transactions that originate from diverse CSs of an SoS can be ordered consistently system wide (i.e., in the total SoS) if a sparse global timestamp identifies every start of a transaction. Such a consistent order is of paramount importance if the transactions access and manipulate the same data resources (e.g., financial account) that are stored redundantly.

The global timestamp added to a transaction makes the transaction *idempotent*. It is much easier to recover from a failure if all involved transactions are idempotent [Mur11].

*C. Synchronization of Output Actions*

In a multi-robot scenario, where different robots, each one controlled by its specific CS, have to interact precisely in the physical space, the global time can be used to accurately synchronize the output actions.

If, because of the jitter of the communication system, the exact instant of arrival of a control message at an actuator node cannot be guaranteed, the concept of a *timed output message* can be used to solve the problem. A *timed output message* contains a timestamp of the future instant when exactly the actuator set-point has to be delivered to the actuator. If a timed output message arrives early at the local controller, the controller will wait such that the set-point will be transmitted at the correct instant.

*D. Conflict Free Resource Allocation*

The resolution of access conflicts to resources that are shared by different CSs can take advantage of the availability of a global time. The widely used TDMA (time-division multiple access) strategy exemplifies this approach.

In a time-triggered communication protocol, the progression of the global time determines when a CS is allowed to send a message on a shared communication medium. The ensuing conflict free-transmission minimizes the transport delay and eliminates the introduction of any communications jitter.

In a periodic control system, where different CSs cooperate to achieve a common control objective, a precise *phase alignment* of processing actions and communication actions that occur in different CSs can be achieved by reference to the global time. The ensuing minimal duration of the resulting control transaction reduces the dead time and improves the quality of control.

*E. Prompt Error Detection*

In Section two we have pointed out that *faults are normal* in an SoS. This characteristic of an SoS has challenging implications for the overall SoS design. An SoS must be structured into well-defined fault-containment units (FCU) and a failure of an FCU must be detected promptly in order that mitigating actions can be performed before the consequences of the failure have propagated to the system level.

One common method to detect fail-silent failures of an FCU is the monitoring of a periodic life sign messages by an error management component. If the instant of sending such a life-sign message and the corresponding time-out instant in the error management component are synchronized by the global time, the *error detection latency* can be minimized. A short error detection latency helps to reduce error propagation effects and increases the probability that the error mitigation will be successful.

The function of shared communication medium can be wiped out if a single babbling node violates the access procedure and continuously tries to send messages. The global time can be used by a communication system to restrict the access to the shared communication resource outside the allocated time interval and to thus rule out the possibility that a failing node can bring down the communication among the correct nodes. Time-triggered communication protocols use this technique to protect the communication system from babbling idiots.

An available global time can also to be used to strengthen security protocols significantly. For example, the validity of keys can be restricted in the temporal domain or replay attacks can be eliminated if the global time is part of every message.

## V. CONCLUSIONS

A global physical time base establishes a universally available temporal framework that can be used in every interface of a CS to solve temporal coordination problems. In this short paper we have discussed the properties of an SoS wide sparse global time-base and shown how such a global time can be used to solve or simplify temporal coordination problems in a System of Systems.

## VI. ACKNOWLEDGEMENTS

This work has been supported, in part, by the European FP7 research project AMADEOS Grant Agreement 610535 on *Systems of Systems*.